\documentclass[aps,prb,showpacs,showkeys,two column,superscriptaddress,preprintnumbers,amsmath,amssymb]{revtex4}

\usepackage{txfonts}
\usepackage[titletoc]{appendix}
\usepackage{amssymb}
\usepackage{array}
\usepackage{mathrsfs}
\newcommand{\PreserveBackslash}[1]{\let\temp=\\#1\let\\=\temp}
\newcolumntype{C}[1]{>{\PreserveBackslash\centering}p{#1}}
\newcolumntype{R}[1]{>{\PreserveBackslash\raggedleft}p{#1}}
\newcolumntype{L}[1]{>{\PreserveBackslash\raggedright}p{#1}}

\usepackage{longtable,bm}
\usepackage{footmisc}
\usepackage{threeparttable}
\usepackage{graphicx,mathrsfs}
\usepackage{dcolumn}
\usepackage{bm,subfigure,overpic}
\usepackage{multirow,booktabs}
\usepackage{color}
\usepackage[]{natbib}
\usepackage{rotfloat}

\begin{document}

\title{Electronic structure and optic absorption of phosphorene under strain}

\author{Houjian Duan}

\author{Mou Yang}
\altaffiliation{Electronic address: yang.mou@hotmail.com}
\author{Ruiqiang Wang}
\affiliation{Laboratory of Quantum Engineering and Quantum Materials, \\
School of Physics and Telecommunication Engineering,
South China Normal University, Guangzhou 510006, China}

\begin{abstract}
We studied the electronic structure and optic absorption of phosphorene (monolayer of black phosphorus) under strain. Strain was found to be a powerful tool for the band structure engineering. The in-plane strain  in armchair or zigzag direction changes the effective masse components along both directions, while the vertical strain only has significant effect on the effective mass in the armchair direction. The band gap is narrowed by compressive in-plane strain and tensile vertical strain. Under certain strain configurations, the gap is closed and the energy band evolutes to the semi-Dirac type: the dispersion is linear in the armchair direction and is gapless quadratic in the zigzag direction. The band-edge optic absorption is completely polarized along the armchair direction, and the polarization rate is reduced when the photon energy increases. Strain not only changes the absorption edge, but also the absorption polarization.
\end{abstract}



\maketitle
\section{introduction}
Exfoliated thin-layer black phosphorus has been realized recently\cite{anisotropic phosphorene}. Phosphorene, a mono-layer of black phosphorus with a finite direct gap, is expected to be a new candidate of the family of pure two-dimensional materials. Phosphorene has attracted widespread interest due to its excellent electronic, optical and mechanical properties. Compared to graphene, phosphorene field-effect transistors have a higher on-off current ratio at room temperature\cite{field-effect transistors, on-off ratio}, which gives phosphorene a great potential for switching device fabrication. Moreover, unlike graphene, which is sensitive to the substrates\cite{substrate 1, substrate 2, substrate 3, substrate 4, substrate 5, substrate 6}, phosphorene is expected to be more reliable. In contrast with transition metal dichalcogenides, phosphorene exhibits a higher carrier mobility.\cite{anisotropic phosphorene, field-effect transistors, on-off ratio, hole mobility, High-mobility and linear dichroism, carrier and gap} Phosphorene has a strongly anisotropic band structure, which allows phosphorene to act as optical polarized-sensitive device\cite{optical properties with thin films,multilayer black phosphorus photodetector}. Due to the puckering lattice structure, phosphorene possesses a superior flexibility and sustains a tensile strain up to about 30$\%$ in either the zigzag or the armchair direction and shows a great power in practical strain engineering.\cite{mechanical flexibility and transition, mechanism and gap} By applying in-plane or vertical strain, the energy gap can be tuned gradually,\cite{strained gap, strained gap and effective mass, strain and electric field with semiconductor-metal transition} and the semiconductor-to-metal transition can be induced.\cite{mechanical flexibility and transition, strain and electric field with semiconductor-metal transition, strained semiconductor-metal transition, gap modification and transition} A considerable increase of effective mass induced by the strain suggests a great potential for switching devices.\cite{strained semiconductor-metal transition, gap modification and transition} Furthermore, strain is regarded as an efficient method to enhance the thermoelectric performance of phosphorene.\cite{thermoelectric properties}

Most literatures about the strain effects on phosphorene are based on first-principle calculations. Numerical methods can handle the complexity of real materials while lack of clear understanding in physics. Recently, tight-banding parameters were obtained from {\it ab initio} calculations, and the dispersion of it fits well with the numerical one.\cite{mono- and bilayer band structure near the band gap} The tight-binding model allows one to get simple solutions to predict various properties of phosphorene under strain and shed more insight on the strain-induced physics.

In this paper, we studied the electronic structure and optic absorption of phosphorene under strain. The band gap depends positively on the in-plane strain and negatively on the vertical strain. By adjusting uniaxial strains in three principle directions, the band gap can be reduced to zero and for this situation, the dispersion is linear in the armchair direction and is quadratic in the zigzag direction. In other words, the semi-Dirac band structure turns up when band gap is closed by strain. A tensile strain along armchair direction increases the effective mass components in both directions, the strain in zigzag direction changes them in two directions in opposite way, and the vertical strain only has significant effect on the effective mass along armchair direction. The band-edge optic absorption is completely polarized along the armchair direction, and the polarization rate decreases when the photon energy increases. Strain changes the absorption polarization as well as absorption edge.

\section{Electronic structure under strain}

\begin{figure}
\includegraphics[width=0.36\textwidth]{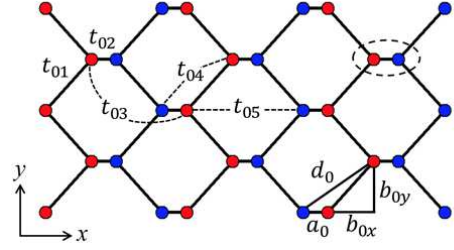}
\caption{(Color Online) The in-plane projection of phosphorene lattice. The red- and blue-filled circles represent the purckled up and purckled down phosphorus atoms. The ellipse denotes the translational cell of the in-plane lattice.}
\label{lattice}
\end{figure}

The in-plane geometry parameters (in units of $\AA$) of the phosphorene lattice are $a_0=0.8014$, $(b_{0x},b_{0y})=(1.515,1.674)$, and the thickness (the distance between sublayers) is $l_0=2.150$.\cite{structure parameter}
The hopping energies $t_{01}$, $t_{02}$, $t_{03}$, $t_{04}$ and $t_{05}$ are $-1.220$, $3.665$, $-0.205$, $-0.105$ and $-0.055$ respectively,\cite{mono- and bilayer band structure near the band gap} all in units of eV. The former two hopping energies play the most significant role in constructing the electronic structure, other smaller ones only modulate the band structure slightly.

When a strain is applied, the lattice mesh is deformed. The deformed coordinates $(x, y, z)$ are related with the undeformed coordinates $(x_0, y_0, z_0)$ by
\begin{eqnarray}\label{tensile model}
\begin{split}
\left(\begin{array}{ccc}
x\\
y\\
z\end{array}\right)=\left(\begin{array}{ccc}
1+\epsilon_x&0&0\\
0&1+\epsilon_y&0\\
0&0&1+\epsilon_z\\
\end{array}\right)\left(\begin{array}{ccc}
x_0\\
y_0\\
z_0\end{array}\right),
\end{split}
\end{eqnarray}
where $\epsilon_i$ is the strain in $i$-direction. In the equation and from here on, we use the subscript $0$ to denote quantities of undeformed phosphorene, and those without the subscript 0 means they are for strained phosphorene. The deformed bond length $r$ can be expressed as
\begin{eqnarray}\label{r}
\begin{split}
r/r_{0}=1+\alpha_x\epsilon_x+\alpha_y\epsilon_y+\alpha_z\epsilon_z.
\end{split}
\end{eqnarray}
where the coefficients $\alpha_i=\partial_{\epsilon_i} r/r_0$ are calculated as
\begin{eqnarray}
\begin{split}
\alpha_x=\frac{x_{0}^2}{r^2_{0}},  \;\;\; \alpha_y=\frac{y_{0}^2}{r^2_{0}}, \;\;\; \alpha_z=&\frac{z_{0}^2}{r^2_{0}}.\\
\end{split}
\end{eqnarray}

The change of the bond length leads to the corresponding modulation of the hopping energy. In phosphorene, the hopping energy is determined by the coupling between $s$ and $p$ orbitals of different phosphorus atoms. Detailed consideration of the coupling reveals that the hopping energy magnitude depends on the bond length in a relation $t\varpropto r^{-2}$.\cite{a relation of hopping energy and bond length 1, a relation of hopping energy and bond length 2} By means of this relation and Eq. (\ref{r}), we have the dependence of hopping energy on the strain components,
\begin{eqnarray}\label{t}
\begin{split}
t/t_0 &= (r/r_0)^{-2}\\
& \approx 1-2\left(\alpha_x\epsilon_x+\alpha_y\epsilon_y+\alpha_z{\epsilon_z}\right).
\end{split}
\end{eqnarray}
In the equation, the second is the linear version of the first line. The linear approximation, which is usually adopted for electronic structure calculation under small strains, is only used for qualitative analysis in this work. The linear coefficients $\alpha_i$ for five strained hopping energies $t_1\sim t_5$ are listed in Tab. \ref{coeffient}.

\begin{table}
\renewcommand\arraystretch{1.5}
\caption{The coefficients $\alpha_i$ for strained hopping energies}
\label{coeffient}
\begin{threeparttable}
\begin{tabular}{cC{1.1cm}C{1.1cm}C{1.8cm}C{1.5cm}C{1.8cm}}
\hline
&$t_{1}$ &$t_{2}$ &$t_{3}$ &$t_{4}$ &$t_{5}$ \\
\hline
$\alpha_x$   &$b^2_{0x} r_{01}^{-2}$    &$a_0^2r_{02}^{-2}$    &$(d_{0x}+a_0)^2 r^{-2}_{03}$   &$d_{0x}^2 r^{-2}_{04}$  &$(d_{0x}+b_{0x})^2r^{-2}_{05}$\\
$\alpha_y$   &$b^2_{0y} r_{01}^{-2}$ &$0$ &$b_{0y}^2 r^{-2}_{03}$ &$b_{0y}^2 r^{-2}_{04}$           &$0$                \\
$\alpha_z$   &0                           &$l_0^2r_{02}^{-2}$     &0                                          &$l_0^2r_{04}^{-2}$                &$l_0^2r_{05}^{-2}$        \\ \hline
\end{tabular}
\end{threeparttable}
\end{table}

Since there is no on-site potential difference of puckered-up and puckered-down atoms, the primitive translational cell consists of two adjacent atoms, as labelled by the ellipse in Fig. \ref{lattice}. The $k$-space Hamiltonian based on the choice of unit cell reads
\begin{eqnarray}\label{H}
\begin{split}
H = & \left(\begin{array}{cc}
g_0 & g_1e^{ik_xa} \\
g^*_1e^{-ik_xa} & g_0 \\
\end{array}\right)
\end{split}
\end{eqnarray}
with
\begin{eqnarray}\label{g}
\begin{split}
g_0=&4t_4{\rm cos} k_x d_{x}\,{\rm cos}k_y d_{y},\\
g_1=&t_2+t_5e^{-i2k_x d_{x}}+2\left(t_1e^{-ik_x d_{x}}+t_3e^{ik_x d_{x}}\right){\rm cos}k_y d_{y}, \nonumber
\end{split}
\end{eqnarray}
where $\boldsymbol{d}=(a+b_{x},b_{y})$. Solving the eigen problem of the Hamiltonian, we have the conduction and valence band energies,
\begin{eqnarray}\label{Ecv}
\begin{split}
E_{c/v}=g_0\pm|g_1|.
\end{split}
\end{eqnarray}

The energy gap, which is the difference between $E_c$ and $E_v$ at $\Gamma$ point, is obtained as
\begin{eqnarray}\label{Eg}
\begin{split}
E_g=2(t_{2}+t_{5}+2t_{1}+2t_{3}).
\end{split}
\end{eqnarray}
The energy gap depends on all hopping energies except for $t_4$, which only accounts for the small asymmetry between the conduction and valence bands. Among these hoppings, $t_2$ and $t_1$ are the largest and second largest ones in amplitude, and have most important influence on the gap. If there is a tensile strain along $z$-direction, $r_2$ is elongated, $t_2$ becomes smaller, and the gap shrinks. If the tensile stain is in $y$-direction, $r_1$ is elongated, $t_1$ is smaller in amplitude, and the gap increases because $t_1$ is negative. When the tensile strain is applied in $x$-direction, both $r_1$ and $r_2$ become longer, which induces opposite effects on the energy gap, so it is difficult to tell how the gap changes. In the linear approximation, the energy gap is
\begin{eqnarray}\label{Eg_linear}
\begin{split}
E_g/E^0_g=1+2.7\epsilon_x+3.8\epsilon_y-8.4\epsilon_z,\\
\end{split}
\end{eqnarray}
The equation verifies the above analysis of $E_g$ when applying $\epsilon_y$ or $\epsilon_z$, and additionally indicates that the gap positively depends on $\epsilon_x$. The energy gap is most sensitive to $\epsilon_z$ and least sensitive to $\epsilon_x$. The dependence of $E_g$ on the in-plane strain and vertical strain in Eq. (\ref{Eg_linear}) agrees with recent literatures\cite{strained gap, strained gap and effective mass}. Interestingly, Eq. (\ref{Eg_linear}) implies that the energy gap can be closed for some strain configurations. For example, zero energy gap can be found at the uniaxial strain $\epsilon_x=-0.37$, $\epsilon_y=-0.26$, or $\epsilon_z=0.12$. The required strain is quite large if the strain is only applied in one direction, but can be lowered by deforming the lattice in three directions simultaneously. For instance, when we applying compressive strains in $x$- and $y$-directions and tensile strain in $z$-direction of the amplitude 0.067, the gap closing can be realized.

From Eq. (\ref{Ecv}), the low-energy dispersions along $x$- and $y$-directions across $\Gamma$ point can be written as
\begin{eqnarray}\label{Ecv_Dirac}
\begin{split}
E_{c/v}(k_x)=& 4t_4\pm\left[{E^2_g} +\left(\gamma_1^2+\gamma_3 E_g\right)d_x^2k_x^2\right]^{1/2},\\
E_{c/v}(k_y)=& 4t_4\pm\left(E_g +\gamma_2d_y^2k_y^2\right),
\end{split}
\end{eqnarray}
with
\begin{eqnarray}
\begin{split}
\gamma_1= &-2\left(t_{1}-t_{3}+t_{5}\right),\\
\gamma_2= &-t_{1}-t_{3},\\
\gamma_3= & -2\left(t_{1}+t_{3}+2t_{5}\right). \nonumber
\end{split}
\end{eqnarray}
The dispersion in $y$-direction is normally parabolic one, and in $x$-direction the dispersion is of the massive Dirac type. When strain is tuned so as to the gap is closed, Eq. (\ref{Ecv_Dirac}) is reduced into
\begin{eqnarray}\label{semi_Dirac}
\begin{split}
E_{c/v}(k_x)=&4t_4\pm \gamma_1d_x k_x,\\
E_{c/v}(k_y)=&4t_4\pm \gamma_2d_y^2k_y^2.\\
\end{split}
\end{eqnarray}
For this case, the dispersion in $x$-direction is reduced to a massless Dirac one with the velocity $\gamma_1d_x/\hbar$, and in $y$-direction the dispersion is gapless parabolic. In other words, the semi-Dirac dispersion\cite{semi-Dirac semi-Weyl semimetal} can be realized when the gap is closed, which can be realized by applying strain.
The low-energy behavior of the semi-Dirac system is extremely anisotropic. The particle kinetics is a mixture of both linear and quadratic dispersions, and which mechanism is dominant depends on the movement orientation.

\begin{figure}
\includegraphics[width=0.47\textwidth]{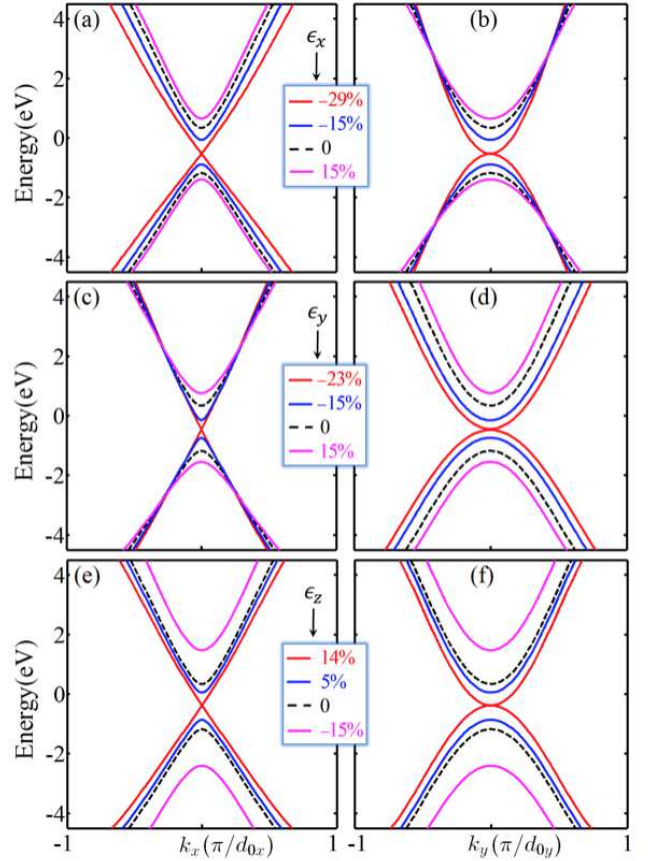}
\caption{(Color Online) Phosphorene dispersions for different uniaxial strains.}
\label{dispersion}
\end{figure}

\begin{figure}
\includegraphics[width=0.47\textwidth]{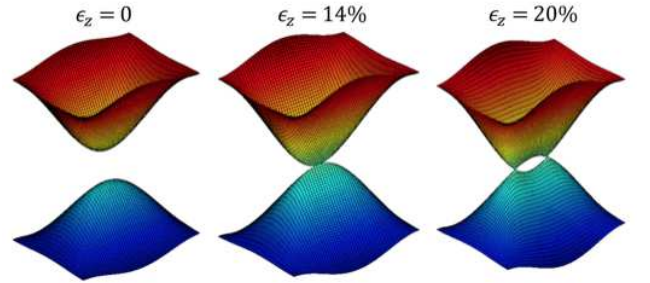}
\caption{(Color Online) The three-dimensional view of normal, semi-Dirac, and band-inversion band structures.}
\label{inversion}
\end{figure}

Figure \ref{dispersion} shows the dispersion evolution when the strain changes. When increasing the compressed in-plane strain or vertical tensile strain, the conduction and valence bands approach to each other. The semi-Dirac band occurs at the uniaxial strain $\epsilon_x=-0.29$, $\epsilon_y=-0.23$, or $\epsilon_z=0.14$, which are qualitatively consistent with what the linear approximation predicts.
If the strain changes on, the conduction band and valence band intersect with each other and the band inversion takes place. Fig. \ref{inversion} illustrates the three-dimensional view of the normal, semi-Dirac, and band-inversion band structures under different strains.

\begin{figure}
\includegraphics[width=0.48\textwidth]{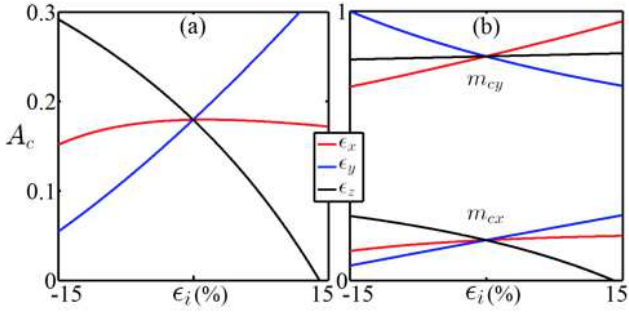}
\caption{(Color Online) (a) The anisotropy ratio and (b) effective mass components of conduction band of phosphorene (in units of  electron mass) as functions of different unaxial strains.}
\label{mass and anisotropy}
\end{figure}

The effective mass component for band $\mu$ in $i$-direction is defined as $\hbar^2 m^{-1}_{\mu i}=|\partial^2E_{\mu}/\partial k^2_i|$ at $\Gamma$ point. According to Eq. (\ref{Ecv_Dirac}), the effective mass along $x$-direction is an explicit function of $E_g$ but that in $y$-direction is not. Strains affect $m_x$ in the similar way as they modify $E_g$, saying, $m_x$ depends on $\epsilon_x$ and $\epsilon_y$ positive but $\epsilon_z$ negatively. By checking the second line of Eq. (\ref{Ecv_Dirac}) and neglecting the non-nearest hoppings $t_3$-$t_5$, the energy dependence on $k_y$ is of the form $t_1d_{y}^2k_y^2$. Because of $t\sim r^{-2}$, we have $m_y$ of the form $r_1^2/r_{1y}^2$. If a tensile strain along $x$-direction is exerted, $r_{1x}$ is elongated and $m_y$ increases; when the strain is along $y$-direction, $r_{1y}$ is stretched and $m_y$ decreases; if the strain is applied vertically, the vertical strain cannot change $r_1$ and $m_y$ keeps unchanged. Figure \ref{mass and anisotropy} (b) shows $m_{cx}$ and $m_{cy}$ as functions of strains, and it coincide with our theoretical analysis.
Phosphorene is a highly anisotropic material, and the effective mass component in the armchair direction is much smaller than that in zigzag direction. The anisotropy of band $\mu$ can be described by the ratio $A_\mu=m_{\mu x}/m_{\mu y}$. The anisotropy ratio of the conduction band as function of strain components can be found in Fig. \ref{mass and anisotropy} (a). $\epsilon_y$ and $\epsilon_z$ have reverse effect on the anisotropy ratio, while $\epsilon_x$ has little influence on it because $\epsilon_x$ changes $m_x$ and $m_y$ synchronously. The valence band are not discussed here because the asymmetry between the conduction and valence bands is quite small.

\par

\section{Optic absorption}
We assume a polarized light irradiates normally on the phosphorene sheet. The light can be described by a time-dependent vector potential ${\bm A}={\bm{\mathcal{E}}}/\omega e^{i\omega \tau}$, where ${\bm{\mathcal{E}}}=(\mathcal{E}_x, \mathcal{E}_y)$ is the electric field vector and $\tau$ is the time. The vector potential is involved into the Hamiltonian by
the Piels substitution ${\bm p}\rightarrow {\bm p}-e{\bm A}$, where ${\bm p}=\hbar\nabla_{\bm k}H/m$ is the momentum operator, and $e$ is the electron charge. According the $A\cdot p$
approximation, the perturbative Hamiltonian caused by the light illumination is $(-e/\hbar\omega) {\bm{\mathcal{E}}}\cdot \nabla_{\bm k}H$. If the photon energy is equivalent to or larger than the band gap, the electrons of the valence band have the probability to be resonantly excited to the conduction band for the condition $E_c-E_v=\hbar\omega$ is met. The absorption rate (number of photons absorbed per unit time) can be calculated by the integration along the equi-energy contour, saying
\begin{eqnarray}\label{I}
\begin{split}
I(\omega) = \frac{e^2}{2\pi\hbar^3\omega^2} \oint_{E_c-E_v=\hbar\omega} |{\bm{\mathcal{E}}}\cdot{\bm{\mathcal{P}}}|^2 \frac{d \boldsymbol{k}}{\nabla_{\boldsymbol k} (E_c-E_v)}.
\end{split}
\end{eqnarray}
with  ${\bm{\mathcal{P}}}=(\mathcal{P}_x,\mathcal{P}_y)$  defined by
\begin{eqnarray}\label{definition}
\begin{split}
\mathcal{P}=\langle\psi_c|\nabla_{\bm k} H|\psi_v\rangle
\end{split}
\end{eqnarray}
where $\psi_c$ and $\psi_v$ are the conduction and valence band states corresponding to the band energies $E_c$ and $E_v$, respectively. In Eq. (\ref{I}) the spin degeneracy is not taken into account.

\begin{figure}
\includegraphics[width=0.46\textwidth]{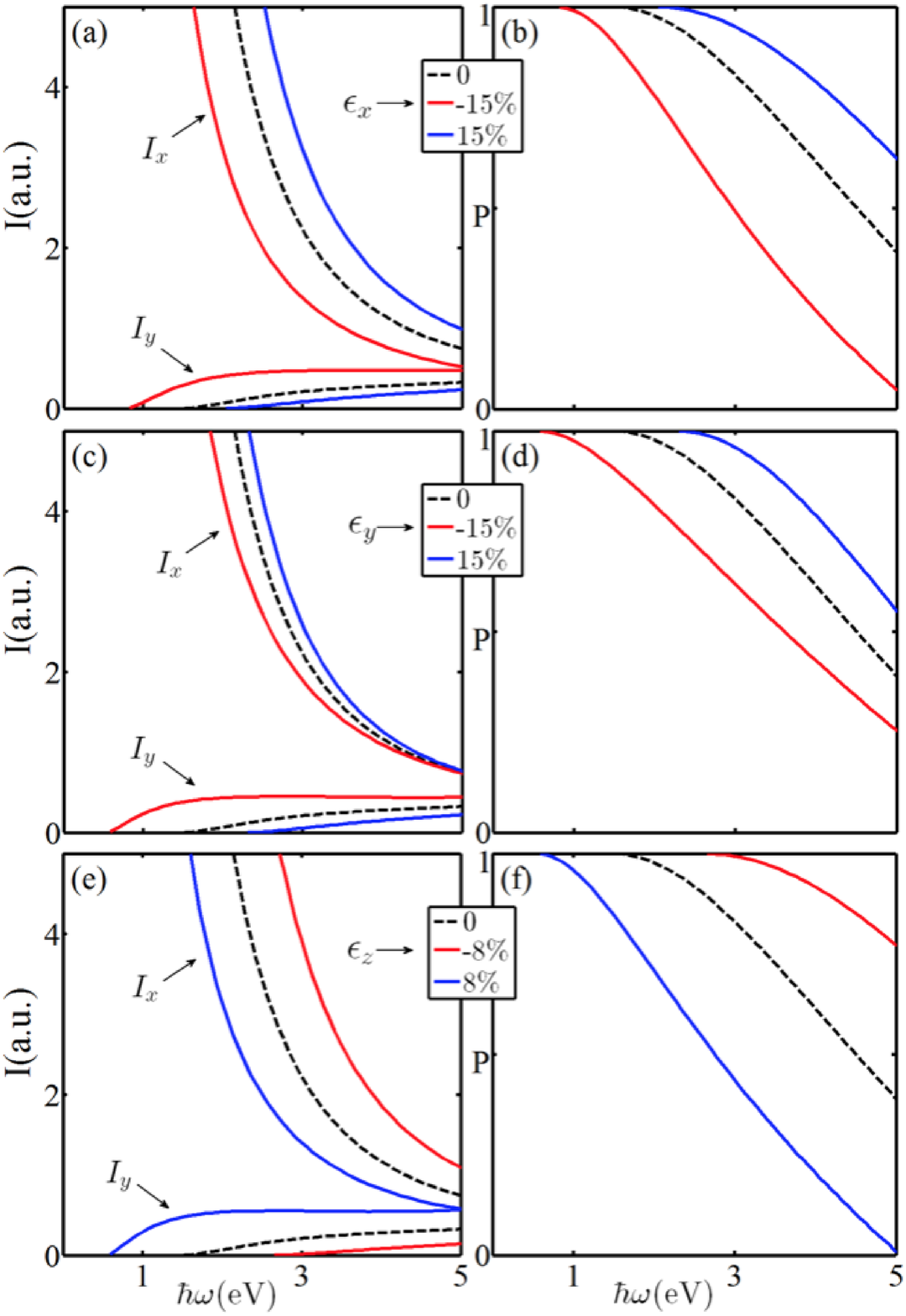}
\caption{(Color Online) Optic absorption rate and its polarization rate under different unaxial strains.}
\label{absorption}
\end{figure}

Figure \ref{absorption} shows the the optic absorption rate for linearly polarized light versus photon energy under different strains.  The $x$-polarized absorption decreases while the $y$-polarized absorption increases from zero on when the photon energy is beyond the band gap. Strain modifies the gap and thus changes the turn-on frequency, and smaller gap leads to more rapidly increasing of $y$-polarized absorption. The absorption for $x$-polarized light is much larger than that for $y$-polarized light and the absorption is highly polarized in the low energy end. At $\Gamma$ point, because of $\partial H/\partial k_y = 0$, we have
\begin{eqnarray}\label{transition element}
\mathcal{P}_y(\Gamma)=&0.
\end{eqnarray}
This means the absorption near the band edge is totally polarized along $x$-direction. When increasing the frequency from the band gap on, the resonant transition occurs at the points deviated from $\Gamma$ point and nonzero $y$-polarized absorption arises. The absorption polarization can be describe by the ratio $P=(I_x-I_y)/(I_x+I_y)$, and it decreases monotonically when the photon energy becomes larger, as shown in Fig \ref{absorption}.

\section{Summary}
We studied the band structure, effective mass, and optic absorption of deformed phosphorene. The energy gap can be decreased by the in-plane or vertical strain. The semi-Dirac dispersion appears when the gap is closed. The effective mass components in the armchair and zigzag directions can be changed by the strain in either direction, but the vertical strain only affects the effective mass along armchair direction. The band-edge absorption is completely polarized along the armchair direction, and the polarization decreases when the light frequency becomes larger.

\acknowledgements
This work was supported by NSF of China Grant No. 11274124,
No. 11474106, and No. 11174088.


\begin{references}
\bibitem{anisotropic phosphorene} F. Xia, H. Wang, and Y. Jia, Nat. Commun. \textbf{5}, 4458 (2014).
\bibitem{field-effect transistors} L. Li, Y. Yu, G. J. Ye, Q. Ge, X. Ou, H. Wu, D. Feng, X. H. Chen, and Y. Zhang, Nat. Nanotechnol. \textbf{9}, 372-377 (2014).
\bibitem{on-off ratio} S. P. Koenig, R. A. Doganov, H. Schmidt, A. H. Neto, and B. ${\rm \ddot{O}}$zyilmaz, Appl. Phys. Lett. \textbf{104}, 103106 (2014).

\bibitem{substrate 1} J. Martin, N. Akerman, G. Ulbricht, T. Lohmann, J. H. Smet, K. von Klitzing, and A. Yacoby, Nature Phys. \textbf{4}, 144-148 (2008).
\bibitem{substrate 2} N. T. Cuong, M. Otani, and S. Okada, Phys. Rev. Lett. \textbf{106}, 106801 (2011).
\bibitem{substrate 3} S. Y. Zhou, G.-H. Gweon, A. V. Fedorov, P. N. First, W. A. de Heer, D.-H. Lee, F. Guinea, A. H. C. Neto, and A. Lanzara, Nature Mater. \textbf{6}, 770-775 (2007).
\bibitem{substrate 4} J. Ristein, S. Mammadov, and T. Seyller, Phys. Rev. Lett. \textbf{108}, 246104 (2012).
\bibitem{substrate 5} G. Giovannetti, P. A. Khomyakov, G. Brocks, V. M. Karpan, J. van den Brink, and P. J. Kelly, Phys. Rev. Lett. \textbf{101}, 026803 (2008).
\bibitem{substrate 6} F. Xia, V. Perebeinos, Y. Lin, Y. Wu, and P. Avouris, Nature Nanotech. \textbf{6}, 179-184 (2011).

\bibitem{hole mobility} H. Liu, A. T. Neal, Z. Zhu, Z. Luo, X. Xu, D. Tom$\mathrm{\acute{a}}$nek, and P. D. Ye, ACS Nano, \textbf{8}, 4033-4041 (2014).
\bibitem{High-mobility and linear dichroism} J. Qiao, X. Kong, Z.-X Hu, F. Yang, and W. Ji, Nat. Commun. \textbf{5}, 4475 (2014).
\bibitem{carrier and gap} Y. Cai, G. Zhang, and Y. Zhang, Sci. Rep. \textbf{4}, 6677 (2014).

\bibitem{optical properties with thin films} T. Low, A. S. Rodin, A. Carvalho, Y. Jiang, H. Wang, F. Xia, and A. H. Castro Neto, Phys. Rev. B \textbf{90}, 075434 (2014).
\bibitem{multilayer black phosphorus photodetector} N. Youngblood, C. Chen, S. J. Koester, and M. Li, Nature Photonics \textbf{10}, 1038 (2015).

\bibitem{mono- and bilayer band structure near the band gap} A. N. Rudenko and M. I. Katsnelson, Phys. Rev. B \textbf{89}, 201408(R) (2014).

\bibitem{mechanical flexibility and transition} Q. Wei, X. Peng, Appl. Phys. Lett. \textbf{104}, 251915 (2014).
\bibitem{mechanism and gap} X. Peng, Q. Wei, A. Copple, Phys. Rev. B \textbf{90}, 085402 (2014).

\bibitem{strained gap} J.-W. Jiang, H. S. Park, Phys. Rev. B \textbf{91}, 235118 (2015).
\bibitem{strained gap and effective mass} X. Han, H. M. Stewart, S. A. Shevlin, C. Richard A. Catlow, Z. X. Guo,  Nano Lett. \textbf{2014}, 14(8).

\bibitem{strain and electric field with semiconductor-metal transition} H. Guo, N. Lu, J. Dai, X. Wu, X. C. Zeng, J. Phys. Chem. C, \textbf{2014}, 118(25).
\bibitem{strained semiconductor-metal transition} M. Elahi, K. Khaliji, S. M. Tabatabaei, M. Pourfath, R. Asgari, Phys. Rev. B \textbf{91}, 115412 (2015).
\bibitem{gap modification and transition} A. S. Rodin, A. Carvalho, A. H. Castro Neto, Phys. Rev. Lett. \textbf{112}, 176801 (2014).

\bibitem{thermoelectric properties} H. Y. Lv, W. J. Lu, D. F. Shao, and Y. P. Sun, Phys. Rev. B \textbf{90}, 085433 (2014).

\bibitem{structure parameter} A. Carvalho, A. S. Rodin and A. H. C. Neto, Europhys. Lett. \textbf{108}, 47005 (2014).

\bibitem{a relation of hopping energy and bond length 1} W. A. Harrison, \emph{Elementary Electronic Structure} (World Scientific, Singapore, 1999).
\bibitem{a relation of hopping energy and bond length 2} H. Tang, J.-W. Jiang, B.-S. Wang, and Z.-B. Su, Solid State Commun. \textbf{149}, 82 (2009).

\bibitem{semi-Dirac semi-Weyl semimetal} S. Banerjee, and W. E. Pickett, Phys. Rev. B \textbf{86}, 075124 (2012).
\end{references}
\end{document}